\documentclass[sigconf, nonacm]{acmart}

\AtBeginDocument{%
  }

\setcopyright{acmcopyright}
\copyrightyear{2022}
\acmYear{2022}
\acmDOI{XXXXXXX.XXXXXXX}

\acmConference[ACM DeFi 2022]{Decentralized Finance and Security}{November 11, 2022}{Los Angeles, CA}
\acmPrice{15.00}
\acmISBN{978-1-4503-XXXX-X/18/06}

\begin{document}

\title{Delta Hedging Liquidity Positions on Automated Market Makers}

\author{Adam Khakhar}
\affiliation{%
  \institution{University of Pennsylvania}
  \country{USA}
}
\email{ak@alumni.upenn.edu}

\author{Xi Chen}
\affiliation{%
  \institution{New York University}
  \country{USA}
}
\email{xc13@stern.nyu.edu}


\begin{abstract}
    Liquidity Providers on Automated Market Makers generate millions of USD in transaction fees daily. However, the net value of a Liquidity Position is vulnerable to price changes in the underlying assets in the pool. The dominant measure of loss in a Liquidity Position is Impermanent Loss. Impermanent Loss for Constant Function Market Makers has been widely studied. We propose a new metric to measure Liquidity Position PNL based on price movement from the underlying assets. Compared to Impermanent Loss, we show how Liquidity Position PNL more appropriately measures the change in the net value of a Liquidity Position as a function of price movement in the assets within the liquidity pool. Our second contribution is an algorithm to delta-hedge arbitrary Liquidity Positions on both uniform liquidity Automated Market Makers (such as Uniswap v2) and concentrated liquidity Automated Market Makers (such as Uniswap v3) via a combination of derivatives.
\end{abstract}

%
%
%




\maketitle

\section{Introduction}
Decentralized Exchanges (DEXs) are a vital component of the Decentralized Finance (DeFi) ecosystem. Decentralized Exchanges are marketplaces where users can trade one cryptocurrency in exchange for another cryptocurrency without giving any group the authority to manage trades or act as a custodian \cite{balancer}. This peer-to-peer trading is accomplished through the use of smart contracts, programmatic agreements fulfilled on the blockchain when predetermined conditions are met \cite{ethorig}. Decentralized Exchanges have several advantages over Centralized Exchanges such as mitigating counter-party risk and reducing the friction to trade due to the lengthy sign-up processes associated with Centralized Exchanges. On a more theoretical level, DEXs can allocate risk among traders with different risk preferences more efficiently, thereby realizing gains from trade that cannot be reproduced in Centralized Exchanges \cite{https://pubs.aeaweb.org/doi/pdfplus/10.1257/aer.20140759}.

Similar to Centralized Exchanges such as Binance, FTX, and Coinbase, Decentralized Exchanges began facilitating trades via Limit Order Books \cite{kpmg-dex}. A Limit Order Book is a method to facilitate an exchange between market participants, where 2 sorted lists are maintained including the price and amount that traders are willing to buy (bid side) or sell (ask side). The Limit Order Book has a unique impact on Order Flow (adding a bid or ask quote to the book) and Trade Flow (accepting a bid or ask price and removing liquidity from the book) \cite{10.2307/2329330}. 0x, dydx, and Serum are examples of Decentralized Exchanges that use a Limit Order Book.

Liquidity is added to Limit Order Books via market participants who add orders to the Limit Book (ex: a bid order where the market participant declares that they are willing to buy $s$ shares at price $p$, or an ask order where the market participant is willing to sell $s$ shares at price $p$). This market participant who adds liquidity to the Limit Order Book is referred to as a Market Maker. Once this order is placed on the limit book, another market participant can accept the full offer or a fraction of the shares in the offer. This market participant who accepts a bid or ask and thereby withdraws liquidity from the Limit Order Book is referred to as the Market Taker. 

In the Limit Order Book paradigm, Market Makers are incentivized to add liquidity through exchanges, which provide benefits for Market Makers such as transaction rebates and reduced transaction fees \cite{binancemm}. In some cases, market participants can agree to become a Contractual Market Maker, where they are compensated to reliably provide liquidity so that the difference between the largest ask and smallest bid is kept to a minimum predetermined range \cite{nysemm}. Low latency trade execution and Market Maker perks have given rise to a substantially profitable set of firms whose strategy is to provide liquidity on the Limit Order Book to capture the Bid-Ask spread while maintaining a delta-neutral portfolio overall \cite{hft}. These firms provide liquidity on the Limit Order Book and profit from Market Maker perks as well as the Bid-Ask spread while having an overall portfolio whose value does not decrease even if the underlying asset changes in price. In the following research, we present the corresponding delta-neutral liquidity provision strategy for Decentralized Exchanges with Automated Market Makers (AMM).

Given the expensive gas fees required to maintain the data structures required for a Limit Order Book on the Ethereum chain, placing and updating orders on the Limit Order Book became prohibitively expensive \cite{gas}. This problem was solved with a new primitive: AMM, which largely replaced the Limit Order Book in most Decentralized Exchanges \cite{kpmg-dex}. In a Decentralized Exchange with an Automated Market Maker, market participants trade against a liquidity pool, with pricing determined using a so-called conservation function \cite{https://doi.org/10.48550/arxiv.2103.12732}. The canonical conservation function is the Constant Function Market Maker, whereby the product of the amounts of each token in the pool are held constant.

Liquidity Providers in an Automated Market Maker provide assets in a pool so that traders can swap one token for another token within this pool. In exchange for providing tokens to a liquidity pool, Liquidity Providers receive a transaction fee from each swap. In Uniswap v3 (the largest Decentralized Exchange by trading volume \cite{theblockresearch}), there are several transaction fee tiers ranging from 1 basis point to 100 basis points \cite{v3whitepaper}.

In 2021, Decentralized Exchanges had over $\$1$ trillion USD in volume traded \cite{theblockresearch}. Since the Summer of 2020, the Uniswap protocol has awarded over $\$1.1$ billion USD in transaction fees to liquidity providers \cite{intotheblock}. However, historically, liquidity providers have lost more in Impermanent Loss than they have received in fees \cite{https://doi.org/10.48550/arxiv.2111.09192}, where Impermanent Loss is defined as the difference between the value of the Liquidity Position and the value of holding initially equal amounts of each asset. The value of a liquidity position is directly dependent on the price of the underlying cryptocurrencies in the pool. Given the large volatility of cryptocurrencies, creating a liquidity position becomes risky \cite{risks9110207}.

Delta-hedging is a strategy which aims to reduce the directional risk associated with price movements of underlying financial instruments within a portfolio \cite{thorpe_kassouf_1967}. In other words, a portfolio is delta-hedged with respect to some assets when the change in price of these assets causes a negligible (or non-existent) change in portfolio value. Many recent works have investigated delta-hedging strategies with desirable properties, such as a delta-hedging strategy via reinforcement learning \cite{dh-rl}.

\section{Motivation}
In Centralized Exchanges such as the New York Stock Exchange, market participants who provide liquidity (add a maker order to the Limit Order Book), have a liquidity position whose portfolio value PNL (profit and loss) is linear with respect to the asset on the exchange. If the liquidity provider is placing an ask order, then the provider is currently holding shares of the asset, which results in a liquidity position portfolio value that is equal to the asset's price multiplied by the number of shares. In this case, the liquidity provider can trivially delta-hedge their position by taking a short position in the same asset. If the liquidity provider is placing a bid order, then this liquidity provider does not own any shares of the asset and the portfolio is trivially delta-neutral with respect to the asset.

Delta-hedging liquidity positions on Automated Market Makers is not as straight forward as it is in the Limit Order Book paradigm. Because the portfolio value of the liquidity position is non-linear with respect to the underlying assets, we must derive a new method to hedge liquidity positions that is tailored to the economics presented in Automated Market Makers. Additionally, the PNL of a liquidity position is more complex than that of a long position on an asset. The PNL of a long position in some asset is equal to the number of shares times the change in price, which trivially has a linear relationship with respect to the asset's price. While market participants can hedge a long position in an asset by taking a short position, there is no reasonable short position that market participants can take on a liquidity position. In other words, there is no market in which users can short a liquidity position.

Given the high volatility of the value of liquidity positions, which have historically generated substantial revenue in transaction fees, we introduce a robust strategy to hedge the value of liquidity positions so that the portfolio value remains nearly constant despite drastic price changes in the underlying assets in the pool. Maintaining a near-constant liquidity position value enables Liquidity Providers to earn transaction fees without risking their liquidity position investment. Our first contribution is a derivation of the true change in value of a liquidity position as a function of price. The dominant measure of loss in a liquidity position is Impermanent Loss; we show why Impermanent Loss is not an appropriate measure of the change in value of a liquidity position as a function of price. Our second contribution is an algorithm to find a combination of derivatives such that the payoff is approximately equal to an arbitrary function of price. Using this algorithm, we can purchase the resulting set of derivatives such that the payoff of the derivatives for any final price plus the change in the value of the liquidity position is equal to 0. This results in a delta-neutral liquidity position. The solution is presented for both AMMs with uniform liquidity and AMMs with concentrated liquidity.

\begin{table}
  \caption{Notation}
  \label{tab:freq}
  \begin{tabular}{ p{8em}  p{14em} }
    \toprule
    Symbol&Definition\\
    \midrule
    $a, b$ & Tokens in a pair of tokens in a liquidity pool, where it is conventional to represent the prices in terms of token $b$. \\
    $p_{a;b}^i, p_{b;b}^i$ & Initial price in units of $b$ upon Liquidity Position entry of token $a$ and token $b$, respectively. \\
    $p_{a;b}^f, p_{b;b}^f$ & Final price in units of $b$ of token $a$ and token $b$, respectively. \\
    $p_{a;b}^l$ & Concentrated liquidity position lower range price in units of $b$ of token $a$. \\
    $p_{a;b}^u$ & Concentrated liquidity position upper range price in units of $b$ of token $a$. \\
    $amount_a^i, amount_b^i$ & Initial total number of tokens in the pool for tokens $a$ and $b$, respectively. \\
    $amount_a^f, amount_b^f$ & Final total number of tokens in the pool for tokens $a$ and $b$, respectively. \\
    $\kappa$ & Constant in Constant Product Formula. \\
    $k$ & Exercise price of an option. \\
    $c$ & Market price of an option. \\
    $\delta$ & Price change from initial price at liquidity provision: $\frac{p_{a;b}^f}{p_{a;b}^i}-1$\\
    PNL & Profit ($\mathbb{R}^{+}$) and loss ($\mathbb{R}^{-}$) from some investment opportunity. \\
  \bottomrule
\end{tabular}
\end{table}

\section{Liquidity Position PNL}
\subsection{Liquidity Position PNL} \label{ssec:LPPNL}
In the following section, we derive a function that measures the PNL of a liquidity position. We wish to hedge the difference between the final value of the liquidity pool assets and the initial value of the liquidity pool assets:
\begin{displaymath}
    \textrm{Liquidity Position PNL} = \frac{\textrm{Final Value of Pool Assets}}{\textrm{Initial Value of LP Investment}}-1
\end{displaymath}

We provide a derivation of Liquidity Position PNL from first principles for Automated Market Makers with uniform liquidity (such as Uniswap v2) and with concentrated liquidity (such as Uniswap v3) in the Appendix Section \ref{deriv-lppnl}. 

\begin{figure}[h]
    \includegraphics[width=9cm]{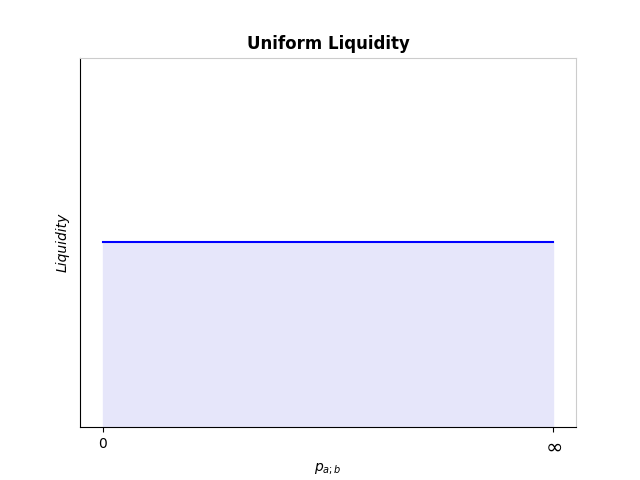}
    \centering
    \caption{Distribution of liquidity over range of price in an AMM with uniform liquidity (Uniswap v2) \cite{v3whitepaper}.}
    \label{fig:uniform-liquidity}
\end{figure}

\begin{figure}[h]
    \includegraphics[width=9cm]{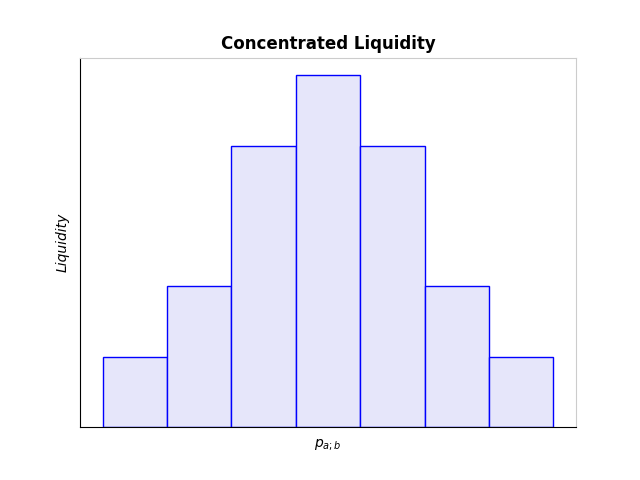}
    \centering
    \caption{Variable distribution of liquidity over range of price in an AMM with concentrated liquidity (Uniswap v3) \cite{v3whitepaper}.}
    \label{fig:concentrated-liquidity}
\end{figure}

We show that for AMMs with uniform liquidity (Appendix Section \ref{ssec:lppnl-ul}):
\begin{displaymath}
    \textrm{Liquidity Position PNL} = \sqrt{\delta + 1}-1
\end{displaymath}

Where $\delta$ represents the price change from the initial price at liquidity provision ($p_{a;b}^i$) to the final price ($p_{a;b}^f$):
$$\delta = \frac{p_{a;b}^f}{p_{a;b}^i}-1$$

\begin{figure}[h]
    \includegraphics[width=10cm]{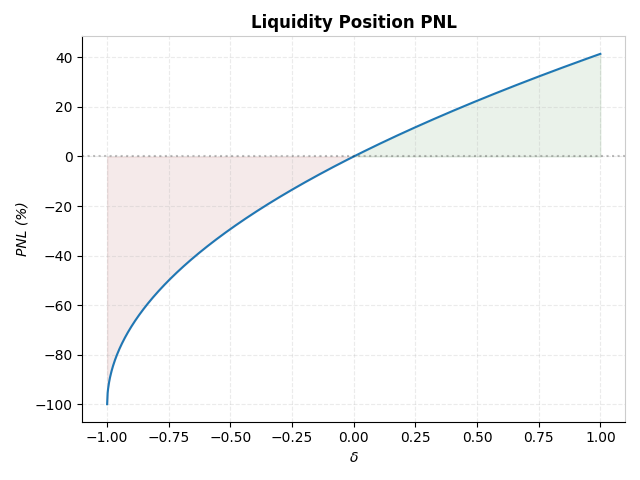}
    \centering
    \caption{Liquidity Position PNL in Uniform Liquidity AMM as a function of change in price ($\delta$).}
    \label{fig:lppnl}
\end{figure}

For AMMs with concentrated liquidity, we show that Liquidity Position PNL is a function of final price, liquidity provision range lower bound, and liquidity provision range upper bound. The proof is provided in the Appendix. In Lemma B: Final Token Amounts in Terms of Price, we find the equation for the final token amounts, which are a function of the final price and liquidity range.
$$\textrm{Liquidity Position PNL} = \frac{p_{a;b}^f \times amount_{a}^f + amount_{b}^f}{p_{a;b}^i \times amount_{a}^i + amount_{b}^i}-1$$
Proof provided in Appendix Section \ref{ssec:lppnl-cl}. Where $amount_a^f, amount_b^f$ are the final amounts of token $a$ and token $b$, respectively:
\begin{multline*} 
    amount_a^f = \\
    \begin{cases}
    \sqrt{\kappa}\cdot \frac{\sqrt{p_{a;b}^u}-\sqrt{p_{a;b}^l}}{\sqrt{p_{a;b}^l \cdot p_{a;b}^u}} & p_{a;b}^f \leq p_{a;b}^l \\
    \sqrt{\kappa}\cdot \frac{\sqrt{p_{a;b}^u}-\sqrt{p_{a;b}^f}}{\sqrt{p_{a;b}^f \cdot p_{a;b}^u}} & p_{a;b}^l < p_{a;b}^f < p_{a;b}^u \\
    0 & p_{a;b}^f \geq p_{a;b}^u
    \end{cases}
\end{multline*}
\begin{multline*}
    amount_b^f = \\
    \begin{cases}
    0 & p_{a;b}^f \leq p_{a;b}^l \\
    \sqrt{\kappa}(\sqrt{p_{a;b}^f}-\sqrt{p_{a;b}^l}) & p_{a;b}^l < p_{a;b}^f < p_{a;b}^u \\
    \sqrt{\kappa}(\sqrt{p_{a;b}^u}-\sqrt{p_{a;b}^l}) \geq p_{a;b}^u
    \end{cases}
\end{multline*}
Proof provided in Appendix Section \ref{lemma-b}, Lemma B: Final Token Amounts in Terms of Price.

\begin{figure}[h]
    \includegraphics[width=9cm]{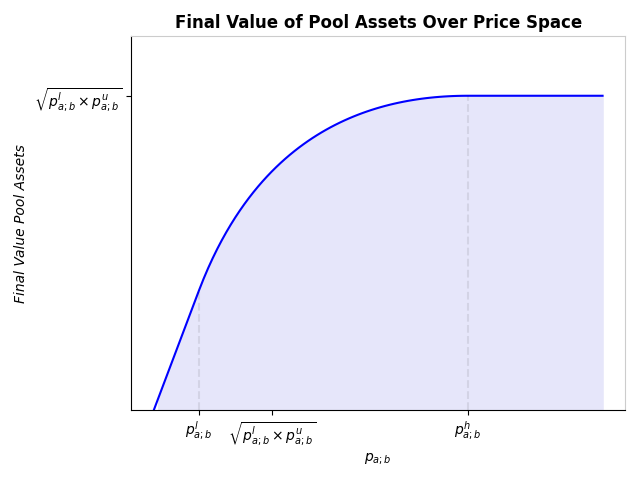}
    \centering
    \caption{Final Value of Pool Assets over Price in an AMM with concentrated liquidity (Uniswap v3) \cite{lambertv3article}.}
    \label{fig:pool_assets_v3}
\end{figure}

\subsection{Discussion of Impermanent Loss}
Impermanent Loss is the most common metric used to analyze the loss incurred in a liquidity position. Impermanent Loss is defined as the difference in portfolio value between a liquidity position and simply holding onto both assets at initially equal values.
$$\textrm{Impermanent Loss} = \frac{\textrm{Final Value of Pool Assets}}{\textrm{Value If Assets Were Held}}-1$$
Impermanent Loss has been well studied on various types of Automated Market Makers \cite{https://doi.org/10.48550/arxiv.2203.11352}.

We provide a derivation of Impermanent Loss from first principles for Automated Market Makers with uniform liquidity (such as Uniswap v2) and with concentrated liquidity (such as Uniswap v3) in the Appendix. We show that for AMMs with uniform liquidity (Appendix Section \ref{ssec:il-ul}):
$$\textrm{Impermanent Loss} = \frac{2\sqrt{\delta+1}}{\delta + 2}-1$$

\begin{figure}[h]
    \includegraphics[width=9cm]{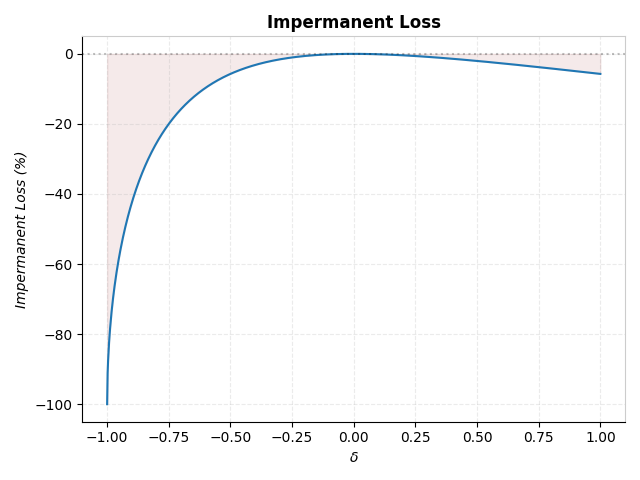}
    \centering
    \caption{Impermanent Loss in Uniform Liquidity AMM as a function of change in price ($\delta$).}
    \label{fig:IL}
\end{figure}

Impermanent Loss and Liquidity Position PNL both share the same numerator: Final Value of Pool Assets:
$$\textrm{Final Value of Pool Assets} = p_{a;b}^f \times amount_{a}^f + p_{b;b}^f \times amount_{b}^f$$
See Figure \ref{fig:pool_assets_v3} for the Final Value of Pool Assets as a function of price. \\

In the case of concentrated liquidity, $amount_a^f$ and $amount_b^f$ depend on the final price as well as the liquidity position range. Thus, similar to Liquidity Position PNL, for AMMs with concentrated liquidity, we show that Impermanent Loss is also function of final price, liquidity provision range lower bound, and liquidity provision range upper bound:
$$\textrm{Impermanent Loss} = \frac{p_{a;b}^f \times amount_{a}^f + amount_{b}^f}{\left( \frac{p_{a;b}^f}{\sqrt{p_{a;b}^i}} + \sqrt{p_{a;b}^i} \right) \times \sqrt{\kappa}}$$
Where Equations (\ref{amt-a-v3}) and (\ref{amt-b-v3}) define $amount_{a}^f$ and $amount_{b}^f$ as functions of price.

Despite the theory surrounding Impermanent Loss, the metric measures the difference between the final value of the liquidity pool assets versus a portfolio of holding an initially equal amount of each asset. As we purchase a portfolio of derivatives, we do not want to hedge for the difference between initially holding an equal amount of each asset. If Impermanent Loss was hedged via derivatives, the resulting portfolio's value would change with directional price movements in the assets held. This would not solve the fundamental problem with liquidity positions: the directional price risk resulting from the high volatility of cryptocurrencies. Furthermore, if we view the liquidity position as a black box function, the market participant invests some combination of tokens into a pool. At any given moment, the investment can be redeemed for some combination of the same tokens in the pool. Because the Liquidity Position PNL metric measures this change in investment value, it is the only necessary value that Liquidity Providers need to hedge. Therefore, unlike Impermanent Loss, Liquidity Position PNL is a direct and efficient metric for measuring the change in value of a Liquidity Position as a function of price.

\section{Using Derivatives to Delta-Hedge}
We define a strategy to be delta-neutral when the directional risk associated with the price movements in the assets held in the trading strategy are nearly completely removed \cite{thorpe_kassouf_1967}. In other words, the change in price of the underlying assets held will not affect the net portfolio value.

\subsection{Derivatives Optimization Problem Definition}
Given the function of Liquidity Position PNL defined in Section \ref{ssec:LPPNL}, which takes as input the final price, we must find some linear combination of derivatives such that:
\begin{displaymath}
    \textrm{Liquidity Position PNL(Final Price)} + \textrm{Payoff(Derivatives, Final Price)} \approx 0
\end{displaymath}

If this equation is satisfied, then any loss in the Liquidity Position resulting from a change in value from the underlying assets will be made up by a positive payoff from the derivatives being held. This results in a trading strategy whose net portfolio value is not affected by directional changes in the price of the underlying assets. The Liquidity Provider can profit from transaction fees without risking losing value in liquidity provided in the pool.

Let $p_{a;b}^f$ be the final price of token $a$ in terms of token $b$, let $c$ be the market price of an option, let $k$ be the strike price of the option. \\
Recall the following equations for the payoff of an option:
\begin{align}
    \textrm{Payoff of Long Call Option} &= max(0, p_{a;b}^f-k) -c \\
    \textrm{Payoff of Short Call Option} &= c - max(0, p_{a;b}^f-k) \\
    \textrm{Payoff of Long Put Option} &= max(0, k-p_{a;b}^f) -c \\
    \textrm{Payoff of Short Put Option} &= c - max(0, k-p_{a;b}^f)
\end{align}

\subsection{Algorithm} \label{ssec:algo}
We can formulate this optimization problem as a least squares regression. We define $\textrm{Payoff(Derivatives, Final Price)}$ as:
\begin{multline}
    \textrm{Payoff(Derivatives, Final Price) } \\ = \sum_{\theta_i \in \theta} \theta_i \times \textrm{Payoff(Derivative}_i\textrm{, Final Price)},
\end{multline}
where $\textrm{Payoff(Derivatives, Final Price)}: \theta, p_{a;b}^f \rightarrow \textrm{PNL}$. \\
We define the cost function of the linear regression as the following:
\begin{multline}
    J(\theta)  = \frac{1}{2}\sum_{p_{a;b}\in P} \Bigl(\textrm{Payoff(Derivatives, Final Price)} \\+ \textrm{Liquidity Position PNL}\Bigr)^2 \nonumber
\end{multline}
Where Liquidity Position PNL is defined in  \ref{ssec:LPPNL}. \\
An additional consideration is the number of options in the portfolio. It is desirable to hold as simple of an options portfolio as possible. It is well known that by adding L1 norm, we can encourage sparsity in $\theta$, thereby resulting in a portfolio with fewer option contracts \cite{tibshirani96regression}.

Adding L1 Norm:
\begin{multline}
    J(\theta)  = \frac{1}{2}\sum_{p_{a;b}\in P} \Bigl( \textrm{Payoff(Derivatives, Final Price)} \nonumber \\ + \textrm{Liquidity Position PNL} \Bigr) ^2  + \lambda \sum_{\theta_i \in \theta} |\theta_i|
\end{multline}

This loss function can be optimized via Stochastic Gradient Descent. Each $\theta_i \in \theta$ can be positive or negative. A positive value is a long position in the options contract, while a negative value is a short position. 

Note that this algorithm identifies a set of options that matches any payoff diagram that can be expressed as a function of $p_{a;b}^f \rightarrow \textrm{PNL}$. In this paper, our goal is to delta-hedge the value of a liquidity position. Thus, in the set of experiments in Section {\ref{experiments}}, we use the algorithm in Section {\ref{ssec:algo}} to find a portfolio of options that negates the Liquidity Position PNL for any $p_{a;b}^f$.

\section{Experimental Results} \label{experiments}
\subsection{Method}
We tested our Delta Hedging algorithm using the Liquidity Position PNL function on both AMMs with uniform liquidity and AMMs with concentrated liquidity. We test to see if the algorithm presented can find a near-optimal set of derivatives for a given Liquidity Position PNL function. In other words, we test the following:

\begin{multline}
\textrm{Liquidity Position PNL(}p_{a;b}^f\textrm{)} \\
 + \textrm{Payoff(Derivative Combination, } p_{a;b}^f \textrm{)}  \\
 \overset{?}{\approx} 0, \quad
 \forall p_{a;b}^f \in (0..\infty) \label{delt-neut}
\end{multline}

Derivatives and their market statistics (bid prices, ask prices, strike price, etc.) are retrieved from the Deribit exchange, which as of July 2022 has $97\%$ Open Interest (O.I.) Market Share of Ethereum Options, $90\%$ O.I. Market Share of Bitcoin Options, and $100\%$ O.I. Market Share of Solana Options \cite{deribit}.

\subsection{Uniform Liquidity Experiment (Uniswap v2)} \label{exp1}
In the following experiment, we tested the effectiveness of the delta-hedging algorithm presented in this paper on the Ethereum-USDT Pair in Uniswap v2, which employs uniform liquidity. \\

Let $Ethereum$ be token $a$ and $USDT$ be token $b$. We execute the experiment with the following steps:
\begin{enumerate}
    \item 
        \begin{sloppypar}
            View transaction for liquidity provision in the Uniswap v2  ETH-USDT pool at address 0x0d4a11d5eeaac28ec3f61d100daf4d40471f1852. Note the price of ETH in terms of USDT - this will be $p_{a;b}^i$
        \end{sloppypar}
    \item Construct a function for Liquidity Position PNL as described in \ref{ssec:LPPNL}.
    \item Retrieve bid and ask quotes for all Ethereum options on Deribit.
    \item Run the algorithm as described in \ref{ssec:algo} to find $\theta$ - the quantities of each option to hold.
    \item Evaluate Equation (\ref{delt-neut})
\end{enumerate}
Deribit had markets for $600$ different Ethereum based options at varying strike prices and exercise dates. This results in $1,200$ possible positions as we can long or short each contract. 

\begin{sloppypar}
    We examine the Liquidity Provision transaction at address 0x863572d128332e1b9ea1fec39b7c0df514af6d67146bb2537c08982c8-6c517b3. From the transaction, we see that $143.78$ $Ethereum$ tokens and $232,015.77$ $USDT$ tokens were provided to the pool. We outline the initial values for variables in Table \ref{tab:init-val-exp-1}.
\end{sloppypar}

\begin{table}
  \caption{Initial Values From Uniform Liquidity Delta Hedging Experiment}
  \label{tab:init-val-exp-1}
  \begin{tabular}{ p{8em}  p{14em} }
    \toprule
    Symbol&Value\\
    \midrule
    $a$& $Ethereum$ \\
    $b$&$USDT$\\
    $p_{a;b}^i$ & $\$1,613.68$ \\
    $p_{b;b}^i$&$\$1$\\
    $amount_a^i$ & $143.78$ \\
    $amount_b^i$&$232,015.77$\\
  \bottomrule
\end{tabular}
\end{table}

\subsubsection{Results From Uniform Liquidity Experiment}
\begin{figure}[h]
    \includegraphics[width=9cm]{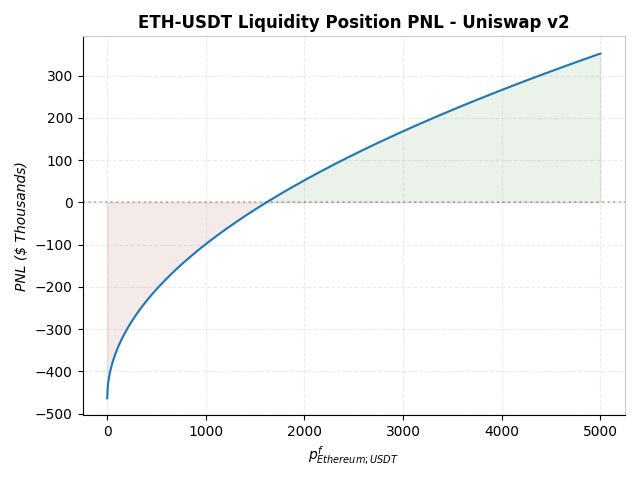}
    \centering
    \caption{Liquidity Position PNL as a function of $p_{a;b}^f$ for Experiment \ref{exp1}.}
    \label{fig:ex1-lppnl}
\end{figure}

Using the Equation in Section \ref{ssec:LPPNL} to find the function for Liquidity Position PNL as a function of final price, we generate Figure \ref{fig:ex1-lppnl}.

We would like to find a portfolio of options that results in the sum of the Liquidity Position PNL and the payoff of the options being equal to 0. Thus, the target PNL for our portfolio of options is found by negating the function for Liquidity Position PNL.

\begin{figure}[h]
    \includegraphics[width=9cm]{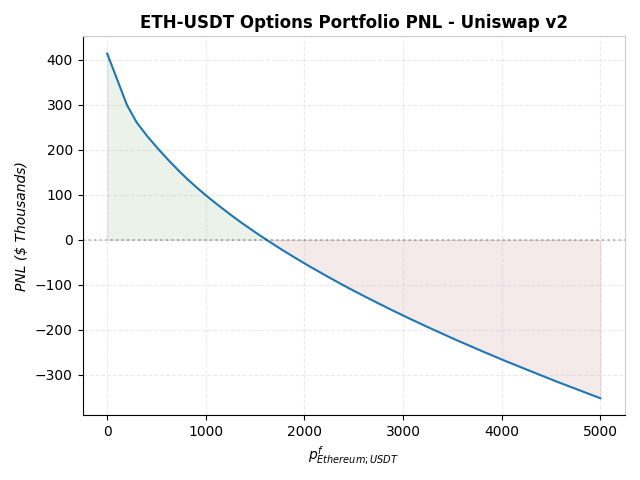}
    \centering
    \caption{Options Portfolio Resulting PNL as a function of $p_{a;b}^f$ for Experiment \ref{exp1}.}
    \label{fig:ex1-optionspnl-actual}
\end{figure}

After running the algorithm described in section \ref{ssec:algo}, we find the following payoff diagram for the constructed portfolio of options: Figure \ref{fig:ex1-optionspnl-actual}.

\begin{figure}[h]
    \includegraphics[width=9cm]{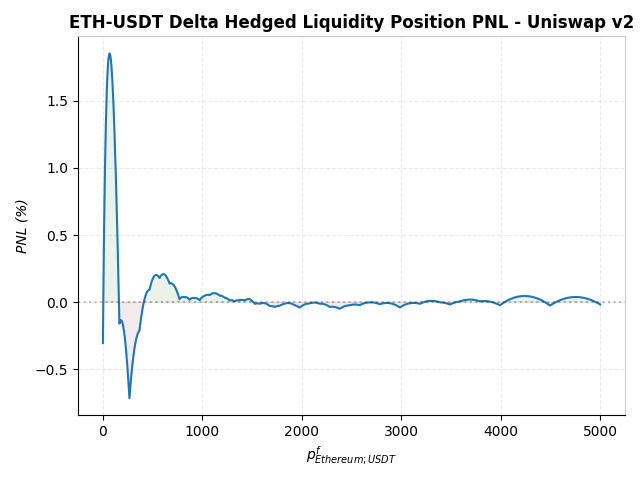}
    \centering
    \caption{Strategy Combined Liquidity Position PNL and Options Portfolio Payoff Diagram for Experiment \ref{exp1}.}
    \label{fig:ex1-stratpnl-actual}
\end{figure}

The following is the resulting total strategy PNL plot: the summation of Liquidity Position PNL and the payoff of the options portfolio Figure \ref{fig:ex1-stratpnl-actual}.

\subsubsection{Uniform Liquidity Experiment Discussion}
From Figure \ref{fig:ex1-stratpnl-actual}, we can see the total strategy PNL as a function of final price in the underlying assets. We can see that the directional risk of price movement in $Ethereum$ is nearly completely removed from the PNL of the combination of the Liquidity Position PNL and portfolio of options. The PNL as a percentage of Liquidity Provided to the pool is nearly $0$ for all final prices of $Ethereum$. We expect to achieve even better performance (where $\text{PNL} \rightarrow 0 \forall p_{a;b}^f$) as the regression model is trained further.

Thus, in the case of uniformly distributed liquidity, the algorithm proposed to generate a portfolio of options to delta-hedge a Liquidity Position is successful. Liquidity Providers can successfully provide liquidity, receive transaction fees from each swap in the pool, and not have directional exposure to $Ethereum$.

\subsection{Concentrated Liquidity Experiment (Uniswap v3)} \label{exp2}
In the following experiment, we tested the effectiveness of the delta-hedging algorithm presented in this paper on the WBTC-USDC Pair in Uniswap v3, which employs concentrated liquidity. $WBTC$ (Wrapped Bitcoin) is an ERC20 token, where each $WBTC$ can be convered into BTC. $WBTC$ enables $BTC$ to be used on the Ethereum chain.\\

Let $Bitcoin$ be token $a$ and $USDC$ be token $b$. We execute the experiment with the following steps:
\begin{enumerate}
    \item 
        \begin{sloppypar}
            View transaction for liquidity provision in the Uniswap v3  WBTC-USDC pool at address 0x99ac8cA7087fA4A2A1FB6357269965A2014ABc35. Note the price of WBTC in terms of USDC - this will be $p_{a;b}^i$
        \end{sloppypar}
    \item Construct a function for Liquidity Position PNL as described in \ref{ssec:LPPNL}.
    \item Retrieve bid and ask quotes for all Bitcoin options on Deribit.
    \item Run the algorithm as described in \ref{ssec:algo} to find $\theta$ - the quantities of each option to hold.
    \item Evaluate Equation (\ref{delt-neut})
\end{enumerate}

\begin{sloppypar}
    We examine the Liquidity Provision transaction at address 0x33b9cf45ce3a3fb36d0c9b2dbaa31b2dd929d10be042b54f7c1ef1852-d44c09a. From the transaction, we see that $19.94$ $WBTC$ tokens and $265,132.51$ $USDC$ tokens were provided to the pool. Additionally, in the transaction logs, we see that the $int24$ values for $tickLower$ and $tickUpper$ are $51960$ and $59940$, respectively. To convert this into a price in human readable format, recall from \cite{v3whitepaper} that:
    \begin{displaymath}
        p_{a;b}=1.0001^{i}
    \end{displaymath}
    Where $i$ is the tick value. Next, note that the number of decimals chosen in the $WBTC$ ERC-20 contract is $8$, while the number of decimals chosen in the $USDC$ ERC-20 contract is 6. Thus, we have to adjust for these order of magnitude differences in order to arrive at the liquidity range upper and lower bounds in units of $USDC$.
    \begin{align*}
        &p_{a;b}^{l}\\
        &=1.0001^{51960} \times 10^{8-6}\\
        &=\$18,050.17
    \end{align*}
    and
    \begin{align*}
        &p_{a;b}^{u}\\
        &=1.0001^{59940} \times 10^{8-6}\\
        &=\$40,089.53
    \end{align*}
    We outline the initial values for variables in Table \ref{tab:init-val-exp-2}.
\end{sloppypar}

\begin{table}
  \caption{Initial Values From Concentrated Liquidity Delta Hedging Experiment}
  \label{tab:init-val-exp-2}
  \begin{tabular}{ p{8em}  p{14em} }
    \toprule
    Symbol&Value\\
    \midrule
    $a$& $WBTC$ \\
    $b$&$USDC$\\
    $p_{a;b}^i$ & $\$23,776.00$ \\
    $p_{b;b}^i$&$\$1$\\
    $amount_a^i$ & $19.94$ \\
    $amount_b^i$&$265,132.51$\\
    $p_{a;b}^l$ & $\$18,050.17$\\
    $p_{a;b}^u$ & $\$40,089.53$\\
  \bottomrule
\end{tabular}
\end{table}

\subsubsection{Results From Concentrated Liquidity Experiment}
\begin{figure}[h]
    \includegraphics[width=9cm]{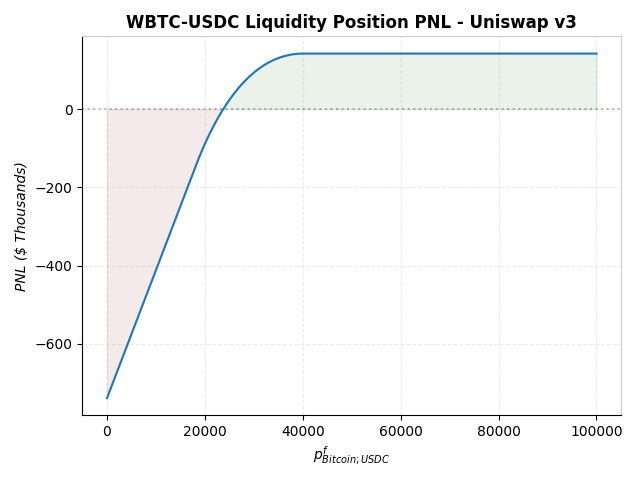}
    \centering
    \caption{Liquidity Position PNL as a function of $p_{a;b}^f$ for Experiment \ref{exp2}.}
    \label{fig:ex2-lppnl}
\end{figure}

Using the Equation in Section \ref{ssec:LPPNL} to find the function for Liquidity Position PNL as a function of final price, we generate Figure \ref{fig:ex2-lppnl}.

We would like to find a portfolio of options that results in the sum of the Liquidity Position PNL and the payoff of the options being equal to 0. Thus, the target PNL function for our portfolio of options is equal to the function for Liquidity Position PNL reflected over the x-axis.

\begin{figure}[h]
    \includegraphics[width=9cm]{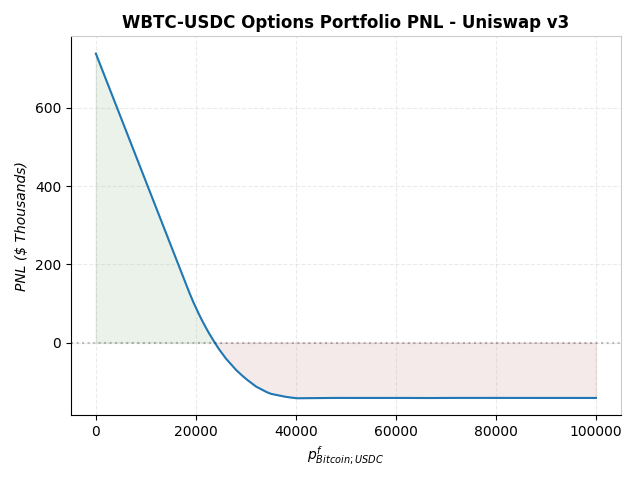}
    \centering
    \caption{Options Portfolio Resulting PNL as a function of $p_{a;b}^f$ for Experiment \ref{exp2}.}
    \label{fig:ex2-optionspnl-actual}
\end{figure}

After running the algorithm described in section \ref{ssec:algo}, we find the following payoff diagram for the constructed portfolio of options: Figure \ref{fig:ex2-optionspnl-actual}.

The following is the resulting total strategy PNL plot: the summation of Liquidity Position PNL and the payoff of the options portfolio Figure \ref{fig:ex1-stratpnl-actual}.
\begin{figure}[h]
    \includegraphics[width=9cm]{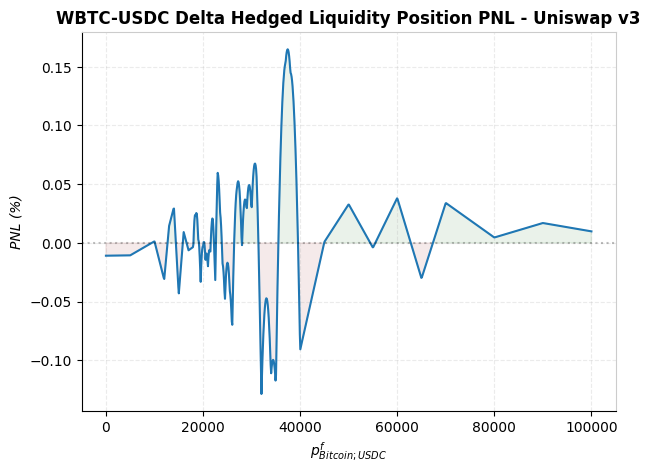}
    \centering
    \caption{Strategy Combined Liquidity Position PNL and Options Portfolio Payoff Diagram for Experiment \ref{exp2}.}
    \label{fig:ex2-stratpnl-actual}
\end{figure}

\subsubsection{Concentrated Liquidity Experiment Discussion}
From Figure \ref{fig:ex2-stratpnl-actual}, we can see the total strategy PNL as a function of final price in the underlying assets. We can see that the directional risk of price movement in $Bitcoin$ is nearly completely removed from the PNL of the combination of the Liquidity Position PNL and portfolio of options. The PNL as a percentage of Liquidity Provided to the pool is $<1\%$ for all final prices of $Bitcoin$.

Thus, in the case of concentrated distributed liquidity, the algorithm proposed to generate a portfolio of options to delta-hedge a Liquidity Position is successful. Liquidity Providers can successfully provide liquidity, receive transaction fees from each swap in the pool, and not have directional exposure to $Bitcoin$.
\bigskip

\section{Discussion}
Our results clearly show that the least squares regression formulation of the problem of finding a near-optimal set of derivatives to approximate the negation of the Liquidity Position PNL function works for Decentralized Exchanges with both uniform liquidity and concentrated liquidity. The value of our approach to delta-hedging liquidity positions lies in its use in the real-world. The delta-hedging strategy described in this paper can be used by any market participant with minimal set-up and algorithm execution costs. Additionally, our delta-hedging strategy requires few trades and does not require a low-latency trading system.

In addition to being used for delta-hedging liquidity positions, the algorithm described in Section \ref{ssec:algo} can be used to find a set of derivatives that approximate any PNL as a function of final asset price. A natural use case is for the replication of the payoff of a liquidity position. In this setting, individuals can participate in the PNL of a liquidity position via purchasing the resulting set of options on a Centralized Exchange such as Deribit, without making any transactions on Decentralized Exchanges.

Loss Versus Rebalancing is another useful metric with many interpretations introduced in concurrent work \cite{lvr}. One method for delta-hedging a liquidity position is achieved by buying and selling the risky asset in the opposite way as the Automated Market Maker does at Centralized Exchange prices \cite{lvr}. While this strategy seems intuitive, it might have potential issues in implementation. Because the strategy requires many taker trades on both the bid and ask side of the Limit Order Book, the market participant will lose value due to the bid-ask spread: the difference between the highest price a buyer is willing to pay (bid side of the Limit Order Book) for the asset and the lowest price a seller is willing to sell the asset (ask side of the Limit Order Book). Even on Binance, the cryptocurrency exchange with the most 24-hour volume, the bid-ask spread can be several basis points (hundredths of a percent) for many cryptocurrencies \cite{topexchange}. If the strategy were to re-balance several hundred times per day, it is easy to see that the portfolio value would quickly diminish due to crossing the bid-ask spread. Comparing the delta-hedging strategy presented in this paper with the one in the Loss Versus Rebalancing work \cite{lvr} would be an interesting direction of future research.

It is important to note the limitations of the options based delta-hedging approach. The model assumes that the derivatives market has sufficient liquidity for the options portfolio. If the options portfolio computed from the algorithm described in Section \ref{ssec:algo} includes a significant position in a specific options contract with little liquidity, the market participant can face high slippage in price when executing the order. Additionally, the payoff expected from the option is only guaranteed when exercising the option. The Deribit options exchange offers European style options, which means that the options can only be exercised at expiry. Therefore, the liquidity provider must decide on an appropriate combination of expiry dates for the options when computing the optimal portfolio.

\section{Conclusion}
In this work, we presented an algorithm for delta-hedging Liquidity Positions via derivatives while introducing a new metric, Liquidity Position PNL, which directly measures the change in the net value of a Liquidity Position as a function of price movement in the underlying assets. The portfolio value of the Liquidity Position combined with the derivatives portfolio determined by the algorithm does not have directional risk associated with price movements in the underlying assets in the liquidity pool. Liquidity Providers can use this algorithm to effectively provide liquidity and earn transaction fees from the protocol while not having to worry about the change in value of the underlying assets in the pool. \\

We are excited about delta-neutral liquidity positions and plan to apply the concept to more tasks such as Strategic Liquidity Provision \cite{https://doi.org/10.48550/arxiv.2106.12033}. We plan to further research quantitative strategies in the economics that Decentralized Exchanges present.\\

We provide a repository including an implementation of the algorithm: \\
\url{https://github.com/adamkhakhar/lp-delta-hedge}

\begin{acks}
We are grateful to Ami Lipkind for their helpful discussions. 
\end{acks}

\bibliographystyle{ACM-Reference-Format}
\bibliography{refs}


\begin{thebibliography}{25}


\ifx \showCODEN    \undefined \def \showCODEN     #1{\unskip}     \fi
\ifx \showDOI      \undefined \def \showDOI       #1{#1}\fi
\ifx \showISBNx    \undefined \def \showISBNx     #1{\unskip}     \fi
\ifx \showISBNxiii \undefined \def \showISBNxiii  #1{\unskip}     \fi
\ifx \showISSN     \undefined \def \showISSN      #1{\unskip}     \fi
\ifx \showLCCN     \undefined \def \showLCCN      #1{\unskip}     \fi
\ifx \shownote     \undefined \def \shownote      #1{#1}          \fi
\ifx \showarticletitle \undefined \def \showarticletitle #1{#1}   \fi
\ifx \showURL      \undefined \def \showURL       {\relax}        \fi
\providecommand\bibfield[2]{#2}
\providecommand\bibinfo[2]{#2}
\providecommand\natexlab[1]{#1}
\providecommand\showeprint[2][]{arXiv:#2}

\bibitem[bin(2022)]%
        {binancemm}
 \bibinfo{year}{2022}\natexlab{}.
\newblock \bibinfo{booktitle}{\emph{Introducing the Binance Market Maker
  Program}}.
\newblock
\urldef\tempurl%
\url{https://www.binance.com/en/support/announcement/360034573691}
\showURL{%
\tempurl}


\bibitem[Best(2022)]%
        {topexchange}
\bibfield{author}{\bibinfo{person}{Raynor~de Best}.}
  \bibinfo{year}{2022}\natexlab{}.
\newblock \bibinfo{title}{Biggest crypto exchanges 2022}.
\newblock
\newblock
\urldef\tempurl%
\url{https://www.statista.com/statistics/864738/leading-cryptocurrency-exchanges-traders/}
\showURL{%
\tempurl}


\bibitem[Biais et~al\mbox{.}(1995)]%
        {10.2307/2329330}
\bibfield{author}{\bibinfo{person}{Bruno Biais}, \bibinfo{person}{Pierre
  Hillion}, {and} \bibinfo{person}{Chester Spatt}.}
  \bibinfo{year}{1995}\natexlab{}.
\newblock \showarticletitle{An Empirical Analysis of the Limit Order Book and
  the Order Flow in the Paris Bourse}.
\newblock \bibinfo{journal}{\emph{The Journal of Finance}}
  \bibinfo{volume}{50}, \bibinfo{number}{5} (\bibinfo{year}{1995}),
  \bibinfo{pages}{1655--1689}.
\newblock
\showISSN{00221082, 15406261}
\urldef\tempurl%
\url{http://www.jstor.org/stable/2329330}
\showURL{%
\tempurl}


\bibitem[Buterin(2014)]%
        {ethorig}
\bibfield{author}{\bibinfo{person}{Vitalik Buterin}.}
  \bibinfo{year}{2014}\natexlab{}.
\newblock \bibinfo{title}{Ethereum: A Next-Generation Smart Contract and
  Decentralized Application Platform}.
\newblock
\newblock
\urldef\tempurl%
\url{https://ethereum.org/669c9e2e2027310b6b3cdce6e1c52962/Ethereum_Whitepaper_-_Buterin_2014.pdf}
\showURL{%
\tempurl}


\bibitem[Cao et~al\mbox{.}(2020)]%
        {dh-rl}
\bibfield{author}{\bibinfo{person}{Jay Cao}, \bibinfo{person}{Jacky Chen},
  \bibinfo{person}{John Hull}, {and} \bibinfo{person}{Zissis Poulos}.}
  \bibinfo{year}{2020}\natexlab{}.
\newblock \showarticletitle{Deep Hedging of Derivatives Using Reinforcement
  Learning}.
\newblock \bibinfo{journal}{\emph{The Journal of Financial Data Science}}
  \bibinfo{volume}{3}, \bibinfo{number}{1} (\bibinfo{date}{dec}
  \bibinfo{year}{2020}), \bibinfo{pages}{10--27}.
\newblock
\urldef\tempurl%
\url{https://doi.org/10.3905/jfds.2020.1.052}
\showDOI{\tempurl}


\bibitem[Cermak et~al\mbox{.}(2022)]%
        {theblockresearch}
\bibfield{author}{\bibinfo{person}{L. Cermak}, \bibinfo{person}{S. Zheng},
  \bibinfo{person}{Andrew Cahill}, \bibinfo{person}{Lars Hoffman}, {and}
  \bibinfo{person}{Eden Au}.} \bibinfo{year}{2022}\natexlab{}.
\newblock \bibinfo{title}{2022 Digital Asset Outlook}.
\newblock
\newblock
\urldef\tempurl%
\url{https://www.tbstat.com/wp/uploads/2021/12/The-Block-Research-2022-Digital-Asset-Outlook.v2.pdf}
\showURL{%
\tempurl}


\bibitem[Doumenis et~al\mbox{.}(2021)]%
        {risks9110207}
\bibfield{author}{\bibinfo{person}{Yianni Doumenis}, \bibinfo{person}{Javad
  Izadi}, \bibinfo{person}{Pradeep Dhamdhere}, \bibinfo{person}{Epameinondas
  Katsikas}, {and} \bibinfo{person}{Dimitrios Koufopoulos}.}
  \bibinfo{year}{2021}\natexlab{}.
\newblock \showarticletitle{A Critical Analysis of Volatility Surprise in
  Bitcoin Cryptocurrency and Other Financial Assets}.
\newblock \bibinfo{journal}{\emph{Risks}} \bibinfo{volume}{9},
  \bibinfo{number}{11} (\bibinfo{year}{2021}).
\newblock
\showISSN{2227-9091}
\urldef\tempurl%
\url{https://doi.org/10.3390/risks9110207}
\showDOI{\tempurl}


\bibitem[Elsts(2021)]%
        {atis}
\bibfield{author}{\bibinfo{person}{Atis Elsts}.}
  \bibinfo{year}{2021}\natexlab{}.
\newblock \bibinfo{title}{Liquidity Math in Uniswap v3}.
\newblock
\newblock
\urldef\tempurl%
\url{https://atiselsts.github.io/pdfs/uniswap-v3-liquidity-math.pdf}
\showURL{%
\tempurl}


\bibitem[Ersan et~al\mbox{.}(2021)]%
        {hft}
\bibfield{author}{\bibinfo{person}{Oguz Ersan}, \bibinfo{person}{Nihan Dalgic},
  \bibinfo{person}{Cumhur Ekinci}, {and} \bibinfo{person}{Mehmet Budur}.}
  \bibinfo{year}{2021}\natexlab{}.
\newblock \bibinfo{title}{High-Frequency Trading and its Impact on Market
  Liquidity: A Review of Literature}.
\newblock
\newblock
\urldef\tempurl%
\url{https://papers.ssrn.com/sol3/papers.cfm?abstract_id=3756596}
\showURL{%
\tempurl}


\bibitem[Faqir-Rhazoui et~al\mbox{.}(2021)]%
        {gas}
\bibfield{author}{\bibinfo{person}{Youssef Faqir-Rhazoui},
  \bibinfo{person}{Javier Arroyo}, \bibinfo{person}{Miller-Janny Ariza-Garzom},
  {and} \bibinfo{person}{Samer Hassan}.} \bibinfo{year}{2021}\natexlab{}.
\newblock \bibinfo{title}{Effect of the Gas Price Surges on User Activity in
  the DAOs of the Ethereum Blockchain}.
\newblock
\newblock
\urldef\tempurl%
\url{https://eprints.ucm.es/id/eprint/64153/1/CHI_21.pdf}
\showURL{%
\tempurl}


\bibitem[{H. Adams et al.}(2021)]%
        {v3whitepaper}
\bibfield{author}{\bibinfo{person}{{H. Adams et al.}}}
  \bibinfo{year}{2021}\natexlab{}.
\newblock \bibinfo{booktitle}{\emph{Uniswap v3 Core}}.
\newblock
\urldef\tempurl%
\url{https://uniswap.org/whitepaper-v3.pdf}
\showURL{%
\tempurl}


\bibitem[Jansen(2022)]%
        {deribit}
\bibfield{author}{\bibinfo{person}{John Jansen}.}
  \bibinfo{year}{2022}\natexlab{}.
\newblock \bibinfo{booktitle}{\emph{Shaping the Crypto Options Industry}}.
\newblock
\urldef\tempurl%
\url{https://www.deribit.com/}
\showURL{%
\tempurl}


\bibitem[Lambert(2021)]%
        {lambertv3article}
\bibfield{author}{\bibinfo{person}{Guillaume Lambert}.}
  \bibinfo{year}{2021}\natexlab{}.
\newblock \bibinfo{booktitle}{\emph{Understanding the Value of Uniswap v3
  Liquidity Positions}}.
\newblock
\urldef\tempurl%
\url{https://lambert-guillaume.medium.com/understanding-the-value-of-uniswap-v3-liquidity-positions-cdaaee127fe7}
\showURL{%
\tempurl}


\bibitem[Loesch et~al\mbox{.}(2021)]%
        {https://doi.org/10.48550/arxiv.2111.09192}
\bibfield{author}{\bibinfo{person}{Stefan Loesch}, \bibinfo{person}{Nate
  Hindman}, \bibinfo{person}{Mark~B Richardson}, {and}
  \bibinfo{person}{Nicholas Welch}.} \bibinfo{year}{2021}\natexlab{}.
\newblock \bibinfo{title}{Impermanent Loss in Uniswap v3}.
\newblock
\newblock
\urldef\tempurl%
\url{https://doi.org/10.48550/ARXIV.2111.09192}
\showDOI{\tempurl}


\bibitem[Malamud and Rostek(2017)]%
        {https://pubs.aeaweb.org/doi/pdfplus/10.1257/aer.20140759}
\bibfield{author}{\bibinfo{person}{Semyon Malamud} {and}
  \bibinfo{person}{Marzena Rostek}.} \bibinfo{year}{2017}\natexlab{}.
\newblock \showarticletitle{Decentralized Exchange}.
\newblock \bibinfo{journal}{\emph{American Economic Review}}
  (\bibinfo{year}{2017}).
\newblock
\urldef\tempurl%
\url{https://pubs.aeaweb.org/doi/pdfplus/10.1257/aer.20140759}
\showURL{%
\tempurl}


\bibitem[Martinelli and Mushegian(2019)]%
        {balancer}
\bibfield{author}{\bibinfo{person}{Fernando Martinelli} {and}
  \bibinfo{person}{Nikolai Mushegian}.} \bibinfo{year}{2019}\natexlab{}.
\newblock \bibinfo{title}{A Non-Custodial Portfolio Manager, Liquidity
  Provider, and Price Sensor}.
\newblock
\newblock
\urldef\tempurl%
\url{https://balancer.fi/whitepaper.pdf}
\showURL{%
\tempurl}


\bibitem[Milionis et~al\mbox{.}(2022)]%
        {lvr}
\bibfield{author}{\bibinfo{person}{Jason Milionis}, \bibinfo{person}{Ciamac~C.
  Moallemi}, \bibinfo{person}{Tim Roughgarden}, {and}
  \bibinfo{person}{Anthony~Lee Zhang}.} \bibinfo{year}{2022}\natexlab{}.
\newblock \bibinfo{title}{Automated Market Making and Loss-Versus-Rebalancing}.
\newblock
\newblock
\urldef\tempurl%
\url{https://doi.org/10.48550/ARXIV.2208.06046}
\showDOI{\tempurl}


\bibitem[Neuder et~al\mbox{.}(2021)]%
        {https://doi.org/10.48550/arxiv.2106.12033}
\bibfield{author}{\bibinfo{person}{Michael Neuder}, \bibinfo{person}{Rithvik
  Rao}, \bibinfo{person}{Daniel~J. Moroz}, {and} \bibinfo{person}{David~C.
  Parkes}.} \bibinfo{year}{2021}\natexlab{}.
\newblock \bibinfo{title}{Strategic Liquidity Provision in Uniswap v3}.
\newblock
\newblock
\urldef\tempurl%
\url{https://doi.org/10.48550/ARXIV.2106.12033}
\showDOI{\tempurl}


\bibitem[Poser(2021)]%
        {nysemm}
\bibfield{author}{\bibinfo{person}{Steven Poser}.}
  \bibinfo{year}{2021}\natexlab{}.
\newblock \bibinfo{title}{Market Makers in Financial Markets: Their Role, How
  They Function, Why They are Important, and the NYSE DMM Difference}.
\newblock
\newblock
\urldef\tempurl%
\url{https://www.nyse.com/publicdocs/nyse/NYSE_Paper_on_Market_Making_Sept_2021.pdf}
\showURL{%
\tempurl}


\bibitem[Robson et~al\mbox{.}(2021)]%
        {kpmg-dex}
\bibfield{author}{\bibinfo{person}{Barnaby Robson}, \bibinfo{person}{Karl
  Koch}, {and} \bibinfo{person}{Arun Ghosh}.} \bibinfo{year}{2021}\natexlab{}.
\newblock \bibinfo{title}{Decentralized Exchanges and Automated Market Makers -
  Innovations, Challenges and Prospects}.
\newblock
\newblock
\urldef\tempurl%
\url{https://assets.kpmg/content/dam/kpmg/cn/pdf/en/2021/10/crypto-insights-part-2-decentralised-exchanges-and-automated-market-makers.pdf}
\showURL{%
\tempurl}


\bibitem[Rodriguez(2022)]%
        {intotheblock}
\bibfield{author}{\bibinfo{person}{Jesús Rodriguez}.}
  \bibinfo{year}{2022}\natexlab{}.
\newblock \bibinfo{booktitle}{\emph{Into the Block: Uniswap Protocol
  Insights}}.
\newblock
\urldef\tempurl%
\url{https://app.intotheblock.com/insights/defi/protocols/uniswap}
\showURL{%
\tempurl}


\bibitem[Thorpe and Kassouf(1967)]%
        {thorpe_kassouf_1967}
\bibfield{author}{\bibinfo{person}{Edward~O. Thorpe} {and}
  \bibinfo{person}{Sheen~T. Kassouf}.} \bibinfo{year}{1967}\natexlab{}.
\newblock \bibinfo{booktitle}{\emph{Beat the Market: A Scientific Stock Market
  System} (\bibinfo{edition}{1st} ed.)}.
\newblock \bibinfo{publisher}{Random House}.
\newblock


\bibitem[Tibshirani(1996)]%
        {tibshirani96regression}
\bibfield{author}{\bibinfo{person}{R. Tibshirani}.}
  \bibinfo{year}{1996}\natexlab{}.
\newblock \showarticletitle{Regression Shrinkage and Selection via the Lasso}.
\newblock \bibinfo{journal}{\emph{Journal of the Royal Statistical Society
  (Series B)}}  \bibinfo{volume}{58} (\bibinfo{year}{1996}),
  \bibinfo{pages}{267--288}.
\newblock


\bibitem[Tiruviluamala et~al\mbox{.}(2022)]%
        {https://doi.org/10.48550/arxiv.2203.11352}
\bibfield{author}{\bibinfo{person}{Neelesh Tiruviluamala},
  \bibinfo{person}{Alexander Port}, {and} \bibinfo{person}{Erik Lewis}.}
  \bibinfo{year}{2022}\natexlab{}.
\newblock \bibinfo{title}{A General Framework for Impermanent Loss in Automated
  Market Makers}.
\newblock
\newblock
\urldef\tempurl%
\url{https://doi.org/10.48550/ARXIV.2203.11352}
\showDOI{\tempurl}


\bibitem[Xu et~al\mbox{.}(2021)]%
        {https://doi.org/10.48550/arxiv.2103.12732}
\bibfield{author}{\bibinfo{person}{Jiahua Xu}, \bibinfo{person}{Krzysztof
  Paruch}, \bibinfo{person}{Simon Cousaert}, {and} \bibinfo{person}{Yebo
  Feng}.} \bibinfo{year}{2021}\natexlab{}.
\newblock \bibinfo{title}{SoK: Decentralized Exchanges (DEX) with Automated
  Market Maker (AMM) Protocols}.
\newblock
\newblock
\urldef\tempurl%
\url{https://doi.org/10.48550/ARXIV.2103.12732}
\showDOI{\tempurl}


\end{thebibliography}

\appendix

\section{Equations}
\begin{itemize}
    \item Note that:
        \begin{equation}\label{pbb}
        p_{b;b}^i=p_{b;b}^f=1
        \end{equation}
    \item Constant Product Formula: 
        \begin{equation}\label{cpf}
            \kappa = amount_{a} \times amount_{b}
        \end{equation}
    \item Price from Constant Product Formula:
        \begin{equation}\label{p}
            p_{a;b} = \frac{amount_b}{amount_a}
        \end{equation}
    \item Let $\delta$ represent the price change from $p_{a;b}^i$ to $p_{a;b}^f$:
        \begin{equation}\label{delta}
            \delta = \frac{p_{a;b}^f}{p_{a;b}^i}-1
        \end{equation}
    \item 
        \begin{equation}\label{il}
            \textrm{Impermanent Loss} = \frac{\textrm{Final Value of Pool Assets}}{\textrm{Value If Assets Were Held}}-1
        \end{equation}
    \item 
        \begin{equation}\label{lppnl}
            \textrm{Liquidity Position PNL} = \frac{\textrm{Final Value of Pool Assets}}{\textrm{Initial Value of LP Investment}}-1
        \end{equation}
\end{itemize}

\section{Lemma A: Relation Between Amount, Price, and $\kappa$}
\begin{align}
    \kappa & = amount_{a} \times amount_{b} && \text{from (\ref{cpf})} \nonumber \\
    \kappa & = \frac{amount_a}{amount_a} \times amount_{a} \times amount_{b} && \text{multiply by } \frac{amount_a}{amount_a} \nonumber \\
    \kappa & = amount_a \times amount_{a} \times p_{a;b} && \text{from (\ref{p})} \nonumber \\
    amount_a &=\sqrt{\frac{\kappa}{p_{a;b}}} && \text{rearrange} \label{amt-a} \\
    amount_b &=\sqrt{\kappa \times {p_{a;b}}} && \text{substitute} \label{amt-b}
\end{align}

\section{Lemma B: Final Token Amounts In Terms of Price} \label{lemma-b}
Given the initial values for price ($p_{a;b}^i$), initial deposit amounts ($amount_a^i, amount_b^i$), liquidity range ($p_{a;b}^l, p_{a;b}^u$), and ending price ($p_{a;b}^f$), we derive a closed-form solution to find the ending token amounts ($amount_a^f, amount_b^f$) \cite{lambertv3article} \cite{https://doi.org/10.48550/arxiv.2111.09192} \cite{atis}.
\subsection{Case $p_{a;b}^f \leq p_{a;b}^l$} \label{ssec:aless}
If the final price of $a$ in terms of $b$ is less than the lower price of the liquidity position range, then only $a$ tokens will remain as all $b$ tokens would be swapped for $a$ tokens (as the price crosses the lower tick). An AMM sells the outperforming asset in exchange for the underperforming asset \cite{https://doi.org/10.48550/arxiv.2111.09192}.
\begin{align*}
    (amount_a + \sqrt{\frac{\kappa}{p_{a;b}^u}})(amount_b + \sqrt{\kappa \cdot p_{a;b}^{l}}) &= \kappa && \text{From \cite{v3whitepaper}} \\
    (amount_a + \sqrt{\frac{\kappa}{p_{a;b}^u}})(\sqrt{\kappa \cdot p_{a;b}^{l}}) &= \kappa && amount_b=0  \\
    amount_a \cdot \sqrt{p_{a;b}^l} + \sqrt{\frac{\kappa \cdot p_{a;b}^l}{p_{a;b}^u}} &= \sqrt{\kappa} && \text{Simplify}
\end{align*}
\begin{align}
    amount_a &= \sqrt{\kappa}\cdot \frac{\sqrt{p_{a;b}^u}-\sqrt{p_{a;b}^l}}{\sqrt{p_{a;b}^l \cdot p_{a;b}^u}} \label{eqn:amt_a}
\end{align}

\subsection{Case $p_{a;b}^f \geq p_{a;b}^u$}
Similar to \ref{ssec:aless}, f the final price of $a$ in terms of $b$ is greater than the upper price of the liquidity position range, then only $b$ tokens will remain as all $a$ tokens would be swapped for $b$ (as the price crosses the upper tick).
\begin{align*}
    (amount_a + \sqrt{\frac{\kappa}{p_{a;b}^u}})(amount_b + \sqrt{\kappa \cdot p_{a;b}^{l}}) &= \kappa && \text{From \cite{v3whitepaper}}  \\
    \sqrt{\frac{\kappa}{p_{a;b}^u}}(amount_b + \sqrt{\kappa \cdot p_{a;b}^{l}}) &= \kappa && amount_a=0   \\
    \frac{amount_b}{\sqrt{p_{a;b}^u}} + \sqrt{\frac{\kappa \cdot p_{a;b}^l}{p_{a;b}^u}} &= \sqrt{\kappa}  \\
\end{align*}
\begin{align}
    amount_b &= \sqrt{\kappa}(\sqrt{p_{a;b}^u}-\sqrt{p_{a;b}^l}) \label{eqn:amt_b} 
\end{align}

\subsection{Case $p_{a;b}^l < p_{a;b}^f < p_{a;b}^u$}
When the ending price is within the liquidity range, we can modify equations (\ref{eqn:amt_a}, \ref{eqn:amt_b}) to account for $p_{a;b}^f$ being within the range. Specifically, for $amount_a$, we can replace $p_{a;b}^l$ with $p_{a;b}^f$ and for $amount_b$, we can replace $p_{a;b}^u$ with $p_{a;b}^f$.
\begin{align}
    amount_a &=\sqrt{\kappa}\cdot \frac{\sqrt{p_{a;b}^u}-\sqrt{p_{a;b}^f}}{\sqrt{p_{a;b}^f \cdot p_{a;b}^u}} \label{eqn:amt_a_within} \\
    amount_b &= \sqrt{\kappa}(\sqrt{p_{a;b}^f}-\sqrt{p_{a;b}^l}) \label{eqn:amt_b_within} 
\end{align}

\subsection{Final Token Amounts in Terms of Price and Liquidity}
\begin{equation} \label{amt-a-v3}
    amount_a^f =
    \begin{cases}
    \sqrt{\kappa}\cdot \frac{\sqrt{p_{a;b}^u}-\sqrt{p_{a;b}^l}}{\sqrt{p_{a;b}^l \cdot p_{a;b}^u}} & p_{a;b}^f \leq p_{a;b}^l \\
    \sqrt{\kappa}\cdot \frac{\sqrt{p_{a;b}^u}-\sqrt{p_{a;b}^f}}{\sqrt{p_{a;b}^f \cdot p_{a;b}^u}} & p_{a;b}^l < p_{a;b}^f < p_{a;b}^u \\
    0 & p_{a;b}^f \geq p_{a;b}^u
    \end{cases}
\end{equation}
\begin{equation} \label{amt-b-v3}
    amount_b^f =
    \begin{cases}
    0 & p_{a;b}^f \leq p_{a;b}^l \\
    \sqrt{\kappa}(\sqrt{p_{a;b}^f}-\sqrt{p_{a;b}^l}) & p_{a;b}^l < p_{a;b}^f < p_{a;b}^u \\
    \sqrt{\kappa}(\sqrt{p_{a;b}^u}-\sqrt{p_{a;b}^l}) \geq p_{a;b}^u
    \end{cases}
\end{equation}

\section{Derivation of Impermanent Loss}
\subsection{Impermanent Loss: Uniform Liquidity} \label{ssec:ilunif}

\subsubsection{Final Value of Pool Assets: Uniform Liquidity} \label{ssec:fvpa}
\begin{align*}
    &\text{Final Value of Pool Assets}\\
    &= p_{a;b}^f \times amount_{a}^f + p_{b;b}^f \times amount_{b}^f \\
    & = p_{a;b}^f \times amount_{a}^f + amount_{b}^f  && \text{from (\ref{pbb})} \\
    & = p_{a;b}^f \times \sqrt{\frac{\kappa}{p_{a;b}^f}} + \sqrt{\kappa \times p_{a;b}^f} && \text{from Lemma A: } (\ref{amt-a}), (\ref{amt-b}) \\
    & = \frac{\sqrt{p_{a;b}^f}}{\sqrt{p_{a;b}^f}} \times \left( p_{a;b}^f \times \sqrt{\frac{\kappa}{p_{a;b}^f}} + \sqrt{\kappa \times p_{a;b}^f} \right) && \text{multiply by } \frac{\sqrt{p_{a;b}^f}}{\sqrt{p_{a;b}^f}} \\
    & = \frac{1}{\sqrt{p_{a;b}^f}} \times\left(  p_{a;b}^f \times \sqrt{\kappa} + p_{a;b}^f \times \sqrt{\kappa} \right) && \text{simplify} \\
    & = 2 \sqrt{\kappa \times p_{a;b}^f} && \text{rearrange}
\end{align*}

\subsubsection{Value if Assets Were Held} \label{ssec:vh}
\begin{align*}
    &\textrm{Value if Assets Were Held} \\
    & = p_{a;b}^f \times amount_{a}^i + p_{b;b}^f \times amount_{b}^i \\
    & = p_{a;b}^f \times amount_{a}^i + amount_{b}^i && \text{from (\ref{pbb})} \\
    & = p_{a;b}^f \times \sqrt{\frac{\kappa}{p_{a;b}^i}} + \sqrt{\kappa \times {p_{a;b}^i}} && \text{from Lemma A: } (\ref{amt-a}), (\ref{amt-b}) \\
    & = \frac{1}{\sqrt{p_{a;b}^i}} \left( p_{a;b}^f \times \sqrt{\kappa} + p_{a;b}^i \times \sqrt{\kappa} \right) && \text{multipy by } \frac{\sqrt{p_{a;b}^i}}{\sqrt{p_{a;b}^i}} \\
    & = \frac{p_{a;b}^f + p_{a;b}^i}{\sqrt{p_{a;b}^i}} \times \sqrt{\kappa} && \text{simplify} \\
    & = \left( \frac{p_{a;b}^f}{\sqrt{p_{a;b}^i}} + \sqrt{p_{a;b}^i} \right) \times \sqrt{\kappa} && \text{simplify} \\
\end{align*}

\subsubsection{Impermanent Loss: Uniform Liquidity} \label{ssec:il-ul}
\begin{align*}
    &\textrm{Impermanent Loss} \\
    & = \frac{\textrm{Final Value of Pool Assets}}{\textrm{Value If Assets Were Held}}-1 && \text{from (\ref{il})} \\
    & = \frac{2 \sqrt{\kappa \times p_{a;b}^f}}{\left( \frac{p_{a;b}^f}{\sqrt{p_{a;b}^i}} + \sqrt{p_{a;b}^i} \right) \times \sqrt{\kappa}}-1 && \text{from \ref{ssec:fvpa} and \ref{ssec:vh}} \\
    & = \frac{2 \sqrt{p_{a;b}^f}}{\left( \frac{p_{a;b}^f}{\sqrt{p_{a;b}^i}} + \sqrt{p_{a;b}^i} \right)}-1 && \text{simplify } \frac{\sqrt{\kappa}}{\sqrt{\kappa}}=1 \\
    & = \frac{2 \sqrt{p_{a;b}^f}}{\sqrt{p_{a;b}^i} \left( \frac{p_{a;b}^f}{p_{a;b}^i} + 1 \right)}-1 && \text{multiply denominator by } \frac{\sqrt{p_{a;b}^i}}{\sqrt{p_{a;b}^i}} \\
    & = \frac{2 \sqrt{p_{a;b}^f}}{\sqrt{p_{a;b}^i}\times (\delta + 2)}-1 && \text{from (\ref{delta})}\\
    & = \frac{2 \sqrt{ \delta + 1}}{\delta + 2}-1 && \text{from (\ref{delta}), } \sqrt{\delta + 1} = \sqrt{\frac{p_{a;b}^f}{p_{a;b}^i}}
\end{align*}

\subsection{Impermanent Loss: Concentrated Liquidity} \label{ssec:ilcl}

\subsubsection{Final Value of Pool Assets: Concentrated Liquidity} \label{ssec:fvpa-cl}
\begin{align*}
    \textrm{Final Value of Pool Assets} &= p_{a;b}^f \times amount_{a}^f + p_{b;b}^f \times amount_{b}^f
\end{align*}
Lemma B, Equations \ref{amt-a-v3} and \ref{amt-b-v3} define $amount_{a}^f$ and $amount_{b}^f$ as functions of price.

\subsubsection{Value if Assets Were Held} \label{ssec:vh-cl}
Same as the case with Uniform Liquidity \ref{ssec:vh}.

\subsubsection{Impermanent Loss: Concentrated Liquidity}
\begin{align*}
    &\textrm{Impermanent Loss} \\
    & = \frac{\textrm{Final Value of Pool Assets}}{\textrm{Value If Assets Were Held}}-1 && \text{from (\ref{il})} \\
    &= \frac{p_{a;b}^f \times amount_{a}^f + amount_{b}^f}{\left( \frac{p_{a;b}^f}{\sqrt{p_{a;b}^i}} + \sqrt{p_{a;b}^i} \right) \times \sqrt{\kappa}} && \text{from \ref{ssec:fvpa-cl}, \ref{ssec:vh-cl}}
\end{align*}
Lemma B, Equations \ref{amt-a-v3} and \ref{amt-b-v3} define $amount_{a}^f$ and $amount_{b}^f$ as functions of price.

\section{Derivation of Liquidity Position PNL} \label{deriv-lppnl}
\subsection{Liquidity Position PNL: Uniform Liquidity} \label{ssec:lppnl-ul}
\subsubsection{Initial Value of LP Investment} \label{ssec:ivlp}
\begin{align*}
    &\textrm{Initial Value of LP Investment} \\
    &= p_{a;b}^i \times amount_{a}^i + p_{b;b}^i \times amount_{b}^i \\
    & = p_{a;b}^i \times amount_{a}^i + amount_{b}^i && \text{from (\ref{pbb})} \\
    & = p_{a;b}^i \times \sqrt{\frac{\kappa}{p_{a;b}^i}} + \sqrt{\kappa \times {p_{a;b}^i}} && \text{from Lemma A: } (\ref{amt-a}), (\ref{amt-b}) \\
    & = 2 \sqrt{\kappa \times {p_{a;b}^i}} && \text{simplify} \\
\end{align*}

\subsubsection{Liquidity Position PNL: Uniform Liquidity}
\begin{align*}
    &\textrm{Liquidity Position PNL} \\
    & = \frac{\textrm{Final Value of Pool Assets}}{\textrm{Initial Value of LP Investment}}-1 && \text{from (\ref{lppnl})} \\
    & = \frac{2 \sqrt{\kappa \times p_{a;b}^f}}{2 \sqrt{\kappa \times {p_{a;b}^i}}}-1 && \text{from \ref{ssec:fvpa} and \ref{ssec:ivlp}} \\
    & = \sqrt{\frac{p_{a;b}^f}{p_{a;b}^i}}-1 && \text{simplify} \\
    & = \sqrt{\delta + 1}-1 && \text{from (\ref{delta}), we have } \frac{p_{a;b}^f}{p_{a;b}^i} = \delta + 1
\end{align*}

\subsection{Liquidity Position PNL: Concentrated Liquidity} \label{ssec:lppnl-cl}
\subsubsection{Initial Value of LP Investment} \label{ssec:ivlp-cl}
Same as the case with Uniform Liquidity \ref{ssec:ivlp}.

\subsubsection{Liquidity Position PNL: Concentrated Liquidity}
\begin{align*}
    &\textrm{Liquidity Position PNL} \\
    & = \frac{\textrm{Final Value of Pool Assets}}{\textrm{Initial Value of LP Investment}}-1 && \text{from (\ref{lppnl})} \\
    & = \frac{\textrm{Final Value of Pool Assets}}{2 \sqrt{\kappa \times {p_{a;b}^i}}}-1 && \text{from \ref{ssec:ivlp-cl}} \\
    & = \frac{p_{a;b}^f \times amount_{a}^f + amount_{b}^f}{2 \sqrt{\kappa \times {p_{a;b}^i}}}-1 && \text{from \ref{ssec:fvpa-cl}} \\
    & = \frac{p_{a;b}^f \times amount_{a}^f + amount_{b}^f}{p_{a;b}^i \times amount_{a}^i + amount_{b}^i}-1 && \text{from \ref{ssec:ivlp-cl}} \\
\end{align*}
Lemma B, Equations \ref{amt-a-v3} and \ref{amt-b-v3} define $amount_{a}^f$ and $amount_{b}^f$ as functions of price.

\end{document}